\documentclass[twocolumn]{revtex4}
\usepackage{graphicx}
\usepackage{epsfig}

\newcommand{\comment}[1]{}

\begin{document}

\title{Replica Field Theory of the Dynamical Transition in Glassy Systems}

\author{Silvio Franz$^1$, Giorgio Parisi$^2$, Federico
  Ricci-Tersenghi$^2$ and Tommaso Rizzo$^2$}

\affiliation{
  $1$ LPTMS, CNRS et Universit\'e Paris-Sud 11 UMR8626, B\^at. 100,
  91405 Orsay Cedex, France\\
  $2$ Dipartimento di Fisica, INFN -- Sezione di Roma I, IPCF-CNR --
  UOS Roma, Sapienza Universit\`a di Roma, P.le Aldo Moro 2, 00185
  Roma, Italy
}

\begin{abstract}
The critical behaviour of the dynamical transition of glassy system is
controlled by a Replica Symmetric action with $n=1$ replicas. The most
divergent diagrams in the loop expansion correspond {\it at all
  orders} to the solutions of a stochastic equation leading to
perturbative dimensional reduction. The theory describe accurately
numerical simulations of mean-field models.
\end{abstract} 

\maketitle

More that twenty years ago it was recognized \cite{KTW} that a certain
class of mean-field spin-glass models (techically speaking, those
where replica symmetry is broken at one step) displays the same
dynamical behaviour predicted by the Mode-Coupling Theory (MCT) of
glasses \cite{MCT}.  This has motivated a great deal of research
\cite{BBRFOT} and many believe that there is an intrinsic analogy
between these two classes of systems.

Lowering the temperature, these models display two transitions. At the
dynamical transition temperature $T_d$ the paramagnetic equilibrium
state abruptly splits into a number of states that is exponentially
large in the size of the system. Correspondingly the equilibrium
dynamics displays the well known MCT phenomenology.  At a static
transition temperature $T_s<T_d$ the number of equilibrium states is
no longer exponential and this is strongly reminiscent of the entropy
crisis which is supposed to happen at the Kauzmann temperature in
glasses.  Between $T_d$ and $T_s$, the entropy crisis supplemented by
nucleation arguments is used to explain the observed super-Arrhenius
slowing down of the dynamics and there are currently many efforts to
verify this scenario.

On the other hand MCT offers a good description of the early stages of
the dynamical slowing down and in this Letter we discuss finite-size
and finite-dimensional effects at $T_d$, i.e.\ at the Mode-Coupling
temperature.  The dynamical transition is characterized by a diverging
correlation length \cite{FPchi4,BB} and therefore it is natural to ask
which are the mean-field critical exponents, what is the upper
critical dimension $d_u$ above which they are valid, and how they are
renormalized below. It is worth noticing that we should rather speak
of pseudo-critical exponents because, due to activated effects, the
dynamical transition is bound to disappear or rather to become a
cross-over in finite dimension, much as in more conventional
metastable phenomena.

In spin-glass models the local variables are usually Ising spins
$s_i=\pm1$ and the relevant order parameter is the overlap between
different equilibrium configurations $q(\sigma,\tau) \equiv N^{-1}
\sum_{i}\sigma_i \tau_i$.  The same order parameter can be studied for
liquids in a lattice gas representation.  It is well known that in
mean-field models the dynamical transition can be located through a
static potential $V(q)$ defined as the average free-energy cost of
imposing that the system stays at an overlap $q$ from a fixed
equilibrium configuration \cite{FP}. The potential has a minimum at
the equilibrium value of the overlap ($q=0$ in the absence of external
fields), but below $T_d$ a secondary minimum appears at a higher
value, $q=q_d$, in the mean-field theory of the problem.  Using the
replica method the potential $V(q)$ can be expressed as a field theory
of an action of a replicated order parameter that is an $n\times n$
overlap matrix $q_{ab}$ with $q_{aa}=0$ (to be continued analytically
to $n\to0$). The paramagnetic solution is $q_{ab}=0$ down to $T_s$,
however at $T_d$ a solution appears with $q_{ab}=q_d$ inside an
$m\times m$ block with $m=1$ (analytic continuation)
\cite{FP,Monasson}.  We want to study the loop expansion around the
latter solution which encodes the presence of an exponential number of
states.  Such an expansion can be simplified noticing that only the
modes with diverging Gaussian fluctuations are relevant for critical
behaviour, while all the others can be integrated out.  Since $T_d$ is
essentially a spinodal point for the $m=1$ solution, it is natural to
assume that the only critical variables $q_{ab}$ are those inside the
$m \times m$ block.  We thus reach the conclusion that {\it the
  relevant field-theory of the problem in the Ginzburg-Landau sense is
  a cubic Replica-Symmetric (RS) field theory with $n=1$ replicas}:
\begin{widetext}
\begin{equation}
{\mathcal L}={1 \over 2}\int dx \left( \sum_{ab} (\nabla
\phi_{ab})^2+m_1 \sum_{ab}\phi_{ab}^2+m_2\sum_{abc}
\phi_{ab}\phi_{ac}+m_3\sum_{abcd}\phi_{ab}\phi_{cd} \right) -{1 \over
  6}\omega_1 \sum_{abc}\phi_{ab}\phi_{bc}\phi_{ca}-{1 \over 6}\omega_2
\sum_{ab}\phi_{ab}^3\;,
\label{T3}
\end{equation}
\end{widetext}
where $\phi_{ab}$ is the difference between the actual value of the
order parameter $q_{ab}$ and its mean-field value $q_d^{MF}$ and
$\phi_{aa}=0$. In the above expression we have retained only the cubic
terms relevant for critical behaviour.  In the critical region we have
$m_1 \propto (T_{d}^{MF}-T)^{1/2}$, where $T_d^{MF}$ is the dynamical
temperature in the mean-field approximation. The corresponding dressed
propagators can be associated to the following physical quantities:
\begin{equation}
G_1(x,y)\equiv \langle
\phi_{ab}(x)\phi_{ab}(y)\rangle_c=\overline{\langle s_x
  s_y\rangle^2}-\overline{\langle s_x\rangle^2}\ \overline{\langle
  s_y\rangle^2}
\label{G1}
\end{equation}
\begin{equation}
G_2(x,y)\equiv \langle
\phi_{ab}(x)\phi_{ac}(y)\rangle_c=\overline{\langle s_x s_y\rangle
  \langle s_x\rangle \langle s_y \rangle}-\overline{\langle
  s_x\rangle^2}\ \overline{\langle s_y\rangle^2}
\label{G2}
\end{equation}
\begin{equation}
G_3(x,y)\equiv \langle \phi_{ab}(x)\phi_{cd}(y)\rangle_c=\overline{
  \langle s_x\rangle^2 \langle s_y\rangle^2}-\overline{\langle
  s_x\rangle^2}\ \overline{\langle s_y\rangle^2}
\label{G3}
\end{equation}
In the r.h.s. of the above expressions the overline means averages
over the quenched disorder {\it and} over the exponential number of
states in which the paramagnetic state splits at $T_d$, while the
various angle brackets mean thermal averages computed inside the {\it
  same} equilibrium state.  We can also consider {\em connected}
correlations w.r.t.\ the thermal noise inside the same state, by
making linear combinations of $G_1$, $G_2$ and $G_3$ that cancel the
last term in eqs.(\ref{G1}--\ref{G3}), e.g.
\begin{eqnarray}
G_{SG}(x,y) &\equiv& G_1-2 G_2 +G_3= \overline{(\langle s_x s_y
  \rangle - \langle s_x \rangle \langle s_y \rangle)^2}\;, \nonumber \\
G_{th}(x,y) &\equiv& G_1-G_2\;,\quad G_q(x,y) \equiv G_1-G_3
\end{eqnarray}
While these {\em connected} correlations describe fluctuations {\it
  inside} a given state, $G_1$, $G_2$ and $G_3$ yield fluctuations
between {\it different} states, and will be called {\em disconnected}
correlations in the following.

We note that the above theory can be used also to describe systems
without quenched disorder, notably {\it structural glasses}. The key
requirement is that there is an exponential number of equilibrium
states and in this case the overline have to be interpreted just as an
average over them. This so-called {\it self-induced} disorder is
supposed to be the bridge between spin-glasses and glasses and it is
the reason why the replica method can be successfully applied to
glasses \cite{MP,SZ}.

At the Gaussian level, fluctuations are controlled by the three
eigenvalues of the quadratic part of $\mathcal{L}$: replicon $r_R$,
longitudinal $r_L$ and anomalous $r_A$ \cite{TDP}. Degeneracies occurs
at special values of $n$: $n=0 \rightarrow r_A=r_L$, $n=1 \rightarrow
r_L=r_R$, $n=2 \rightarrow r_A=r_R$.  We have found that, as a
consequence of the degeneracy at $n=1$ between the replicon and
longitudinal eigenvalues, a double pole appears in the bare
propagators. Switching to momentum representation the leading
divergent behaviour is given by:
\begin{eqnarray}
G_1(k) & = & {2 \, a \over (k^2+r_R)^2}-{1 \over k^2+r_R}+ {2 \over
  r_A} \nonumber\; , \\ G_2(k) & = & {2 \, a \over (k^2+r_R)^2}-{3 \over
  k^2+r_R}+ {3 \over r_A} \nonumber\; , \\ G_3(k) & = & {2 \, a \over
  (k^2+r_R)^2}-{4 \over k^2+r_R}+ {4 \over r_A} \; ,\nonumber
\end{eqnarray}
where $r_R=r_L=2 m_1$ and $a \equiv \lim_{n \rightarrow 1}(r_R-r_L) /
(n-1)=-2m_2-2m_3$. Note that $r_A$ is not critical and remains
non-zero.  The presence of an unexpected double pole is similar to
what happens in the Random Field Ising Model (RFIM) \cite{DDG}.
Another feature of the above expressions that resembles the RFIM is
the fact that any {\it connected} correlation diverges instead with a
single pole.  It is well known that the perturbative loop expansion of
the RFIM is the same of a stochastic equation \cite{PS1}: quite
surprisingly we have found that this property is also shared by the RS
field theory with $n=1$. We stress that in the literature similar loop
expansions are usually limited to the first few orders \cite{DDG,TDP}
for general values of $n$ because for each diagram one has to perform
a complex summation over replica indices.  Therefore it is remarkable
that in the case $n=1$ one controls the loop expansion at all
orders. The most divergent diagrams in the loop expansion of the
theory (\ref{T3}) corresponds {\it at all orders} to the solution of
the following cubic equation in presence of a quenched Gaussian random
field $h(x)$ \footnote{The result can be derived either
  diagramatically or directly following the approach of Cardy
  \cite{Cardy} for the RFIM model.}
\begin{eqnarray}
-\Delta \phi+r_R \phi+\omega \phi^2 & = & h \; ,\label{stocha}\\
\text{with} \quad\quad [ h(x)h(y) ]_h & = & 2 \, a \, \delta(x-y)\;.
\nonumber
\end{eqnarray}
There is an unique cubic constant $\omega=\omega_2-\omega_1$ and
$[ \dots ]_h$ means average over the random field.  The precise
meaning of the equivalence is that the most divergent diagrams in the
loop expansion of $G_1(x,y)$, $G_2(x,y)$, $G_3(x,y)$ coincide to all
orders with the loop expansion of $[\phi(x)\phi(y)]_h$ where $\phi$ is
solution to eq.~(\ref{stocha}).  The correspondence also holds at the
level of the less divergent {\it connected} correlations functions and
we have:
\begin{equation}
G_{SG}=[\phi h ]_h\,,\;G_{th}=2 [\phi h ]_h\,,\;G_q=3[\phi h]_h
\end{equation}
The stochastic equation (\ref{stocha}) leads to a simple physical
interpretation: {\it critical behavior is controlled by the random
  field fluctuations from state-to-state}.  The diagrammatic analisys
shows that the upper critical dimension is $d_u=8$ and not the naive
expectation 6. Furthermore the mapping to the stochastic equation
implies a perturbative dimensional reduction \cite{PS1}, suggesting
that the critical exponents are the same of the pure model in
dimension $D-2$. In the RFIM dimensional reduction does not hold
because of non perturbative effects \cite{PS1} while it does hold for
branched polymers \cite{PS2}. There is no general recipe to know if it
holds or not and we will not further comment on this point.

At the mean-field level the theory predicts critical exponents
different from those of a standard cubic theory. Remarkably the
predictions of the latter have been found to disagree with numerical
simulations in a recent study \cite{SBBB}.  In mean-field it is
natural to study the critical behavior of integrated quantities
diverging at the critical point, like the susceptibilities
$\chi_\bullet \equiv N^{-1} \sum_{x,y} G_\bullet(x,y)$, where $N$ is
the system size.  As usual the loop expansion can be recast formally
in order to deal with divergences at criticality, and the result is
that disconnected susceptibilities diverge as $N^{1/2}$, while
connected susceptibilities diverge as $N^{1/4}$. However the
prefactors are expressed as series with all positive coefficients, and
are not resummable. This can be seen also noticing that the stochastic
equation (\ref{stocha}) does not admit a real solution for a
sufficiently negative field meaning that the averages $[\dots ]_h$ are
not well defined beyond perturbation theory. This is precisely what we
were expecting because we are dealing with metastable states that are
intrinsically ill-defined at the critical temperature. On the other
hand, dynamics is always well defined, and allow us to access the
critical region (e.g.\ in numerical simulations) to test the above
critical exponents.

We have studied numerically the dynamics at $T_d$, using a mean-field
model of $N$ Ising spins $s_i=\pm1$, interacting by 3-spin couplings
($J_{ijk}=\pm1$), randomly chosen such that each spin participates
exactly to $z=8$ interactions. Starting from an equilibrated
configuration, we measured the overlap
$C(t)=\sum_{i=1}^Ns_i(0)s_i(t)/N$ between the initial configuration
and the configuration at time $t$. Below $T_d$ the initial condition
determines the state in which the system will remain trapped along the
dynamics.  The fluctuations of $C(t)$ are called $\chi_4(t)$ in the
glass literature \cite{FPchi4}. However, according to our results it
is crucial to distinguish between {\it disconnected} and {\it
  connected} fluctuations; so we write $\chi_4(t) = \chi_{het}(t) +
\chi_{th}(t)$ with
\begin{eqnarray}
\chi_{het}(t) &\equiv& N \left(\overline{\langle
  C(t)\rangle^2}-\overline{\langle C(t) \rangle}^2\right)\;,\\
\chi_{th}(t) &\equiv& N \;\overline{\langle C(t)^2\rangle-\langle C(t)
  \rangle^2}\;.
\end{eqnarray}
In the $t\to\infty$ limit, we have that $\chi_{het}(t)\to\chi_2 \equiv
N^{-1}\sum_{x,y}G_2(x,y)$, $\chi_{th}(t)\to\chi_{th} \equiv
N^{-1}\sum_{x,y}G_{th}(x,y)$ and $\chi_4(t)\to\chi_1 \equiv
N^{-1}\sum_{x,y}G_1(x,y)$. This connection allow us to test the above
scaling predictions through dynamical measurements.

Numerically we work on the Nishimori line, such that the starting
configuration with all $s_i(0)=1$ is an equilibrium configuration
\cite{NishiInit}. Therefore for each sample we consider just one of
the exponentially many typical states and averages between different
states are obtained changing the sample.  Averages are computed by a
number of samples (states) $N_S$ such that $N N_S=3 \cdot 10^7$ and
with 2 real replicas per sample evolving with different thermal
noises. The dynamics exhibits the MCT phenomenology. In particular
above $T_d$ the average correlation $C_{av}(t) \equiv
\overline{\langle C(t) \rangle}$ relaxes to zero in a two-step
fashion. On the time scale $t_{\beta}$ of the $\beta$ regime
$C_{av}(t)$ remains around a plateau value $C_p$, while one the larger
time scale $t_{\alpha}$ of the $\alpha$ regime it decays to zero. Both
the two time-scales diverge at $T_d$ and the system remains in the
state selected by the initial configuration. The analytical solution
of the model \cite{MRT} gives $T_d=1.3420(5)$ and $C_p=0.750(5)$.
Even at $T=T_d$, due to finite size effects, $C_{av}(t)$ eventually
decays to zero although on time scales diverging with the system
size. In present context we do not want to discuss the behaviour of
these time scales and we find convenient to consider reparameterized
quantities $\chi_{het}(C_{av})$ and $\chi_{th}(C_{av})$ \cite{HKproc}.
We may distinguish three regions.

{\it The perturbative region} corresponds to $C_{av}>C_p$. It is
reached in times of order $O(1)$, such that fluctuations are themself
$O(1)$ and admit a regular expansion in powers of $1/N$. Using rather
natural finite-time-scaling arguments one can match these dynamical
perturbative series with the perturbative series of the statics
generated by action (\ref{T3}). It can be argued that the coefficients
of the various terms in the expansions {\it diverge} with the same
power of $\delta C \equiv (C_{av}-C_p)$ of the corresponding static
quantity.  In particular the most divergent terms at any order are
given by
\begin{eqnarray}
\chi_{het}(C_p + \delta C)& = & {1 \over \delta
  C^2}\sum_{k=0}^\infty{c_k \over (N \delta C^4)^k}\;,
\label{GCC1} \\
\chi_{th}(C_p + \delta C) & = & {1 \over \delta
  C}\sum_{k=0}^\infty{d_k \over (N \delta C^4)^k} \;.
\label{GCC2}
\end{eqnarray}

\begin{figure}[t]
\begin{center}
\epsfig{file=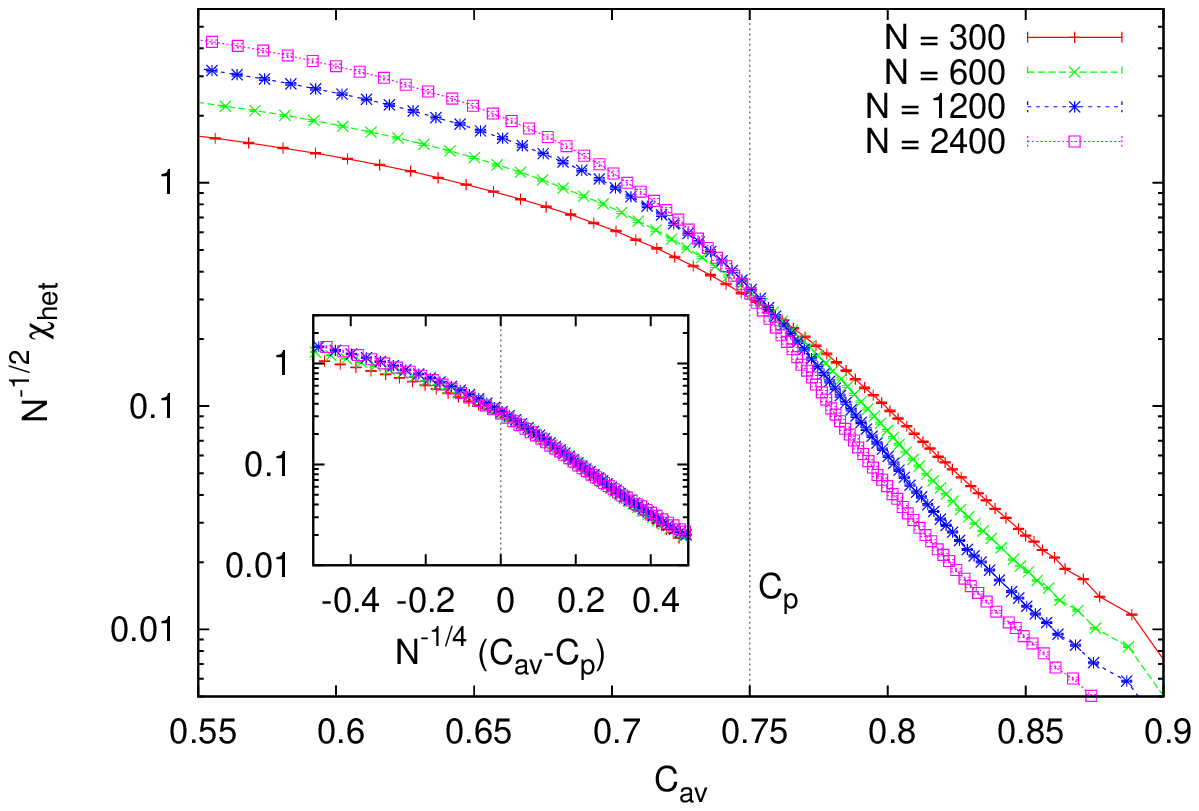,width=\columnwidth}
\epsfig{file=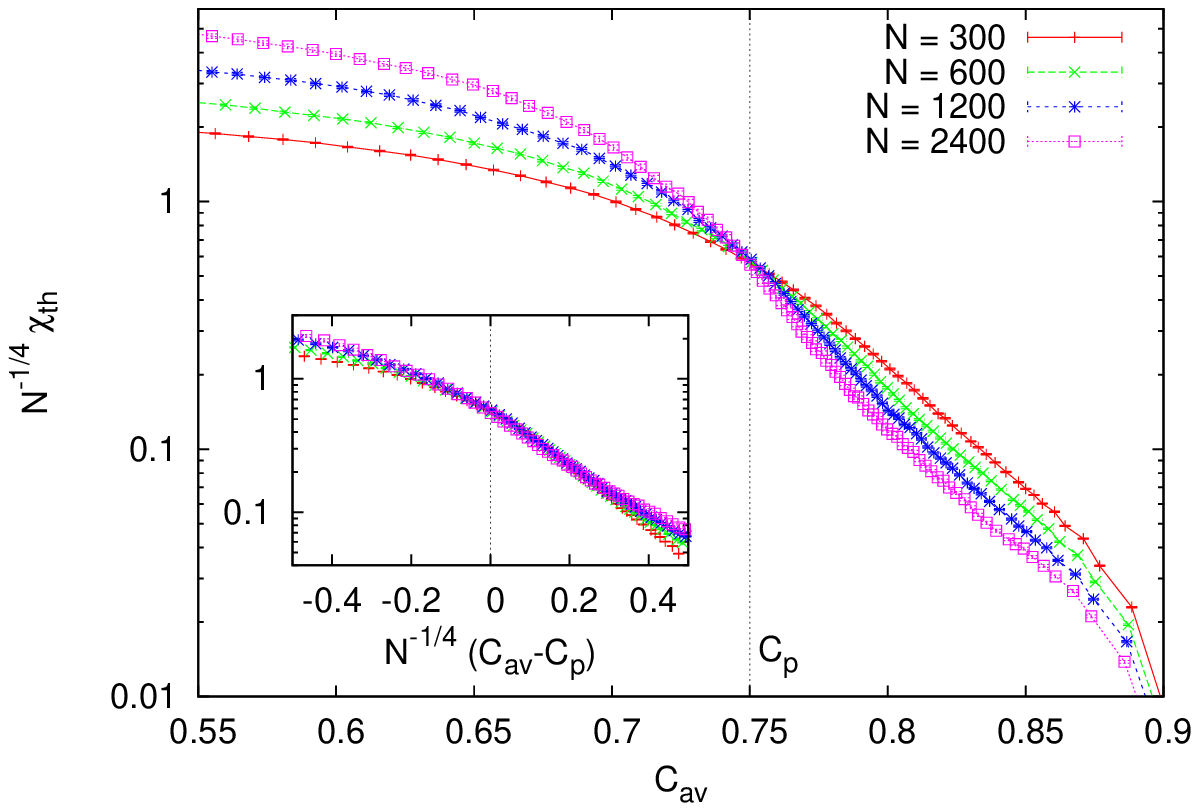,width=\columnwidth}
\caption{(color online) Rescaled disconnected susceptibility $N^{-1/2}
  \chi_{het}(C_{av})$ (top) and rescaled connected susceptibility
  $N^{-1/4} \chi_{th}(C_{av})$ (bottom). The curves for different
  system sizes cross at the analytically known plateau value
  $C_p=0.75$. The insets show the same data rescaled horizontally
  according to the scaling variable $x=N^{-1/4}(C_{av}-C_p)$, leading
  to a very good data collapse.}
\label{C4}
\end{center}
\end{figure}

{\it The scaling region} corresponds to values of $C_{av}(t)$ near
$C_p$. This region is explored on a time scale diverging with the
system size and we expect diverging fluctuations. According to the
above perturbative series the scaling region corresponds to $\delta
C=O(N^{-1/4})$, with the following scaling laws
\begin{eqnarray}
\chi_{het}(C_p + N ^{-1/4} x) &=& N^{1/2} f_{het}(x)\;, \\
\chi_{th}(C_p + N ^{-1/4} x) &=& N^{1/4} f_{th}(x)\;,
\label{scaling}
\end{eqnarray} 
where the two scaling functions $f_{het}(x)$ and $f_{th}(x)$ go to
zero for $x \rightarrow +\infty$ and diverge for $x \rightarrow
-\infty$. The above perturbative series provide their asymptotic
expansion for $x \rightarrow \infty$: at leading order $f_{het}(x)
\sim x^{-2}$ and $f_{th}(x) \sim x^{-1}$. The numerical data are in
very good agreement with the expected behaviour, see Fig.~\ref{C4}.

\begin{figure}[t]
\begin{center}
\epsfig{file=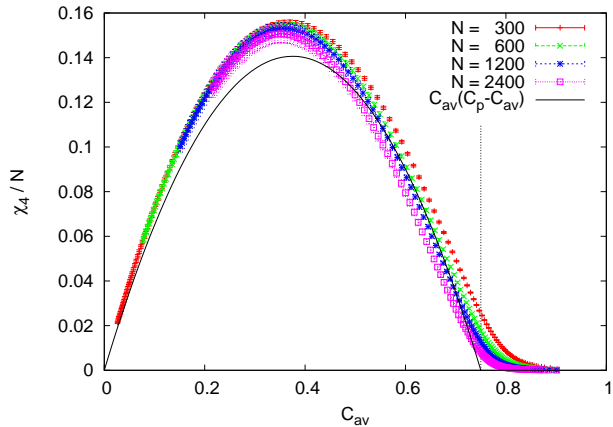,width=\columnwidth}
\caption{Normalized disconnected susceptibility $\chi_4$ in the
  $\alpha$ regime, $C_{av}<C_p$. }
\label{ALFA}
\end{center}
\end{figure}

{\it The $\alpha$ regime} corresponds to $0<C_{av}<C_p$ and we can not
use any perturbative information here. In this region our numerical
data are close to the form $\chi_4(C_{av}) = N C_{av} (C_{p}-C_{av})$,
see Fig.~\ref{ALFA}, which would hold if, for any initial
configuration and thermal noise, the relaxation consisted in a sharp
jump from the plateau value $C_p$ to the value for uncorrelated
configurations, $C=0$.

It is worth noticing that it is far from trivial that the exponents
predicted in perturbation theory can be actually observed in the
critical region. Indeed if non-perturbative effects were also present
on the same time scale they would wash out the perturbative
results. Our results suggest that non-perturbative effects appear
instead on a larger time scale. The time scale of the critical region
diverges with the system size as $N^{1 \over 4a}$ and a reasonable
possibility is that the time scale of the $\alpha$ regime scales like
$N^{{1 \over 4a}+{1 \over 4b}}$, where $a$ and $b$ are the dynamical
MCT exponents of the $\beta$ regime.  The results of the static theory
can be used to safely infer other properties of critical dynamics in
the early $\beta$ regime.  In particular, for $\tau \equiv T-T_d > 0$,
we have a $\beta$-like regime for short time, $t < t_{\beta} =
O(\tau^{-1/2a})$, where the following scaling laws should hold
\begin{equation}
\chi_{het}(t) = \frac{1}{\tau}\, g_{het}(t/t_{\beta})\;, \quad
\chi_{th}(t) = \frac{1}{\sqrt{\tau}}\, g_{th}(t/t_{\beta})\;.
\end{equation}
Correspondingly, we have $\chi_4(t) \propto \chi_{het}(t) \propto
t^{2a}$ and $\chi_{th}(t) \propto t^a$ at finite times.  It could be
that one is able by means of matching arguments and numerical
observation to access also the late $\beta$ regime and the $\alpha$
regime. We choose however not to discuss this point because we think
that a satisfactory understanding of the $\alpha$ regime shall include
an analitycal treatment of non-perturbative effects.


\begin{thebibliography}{99}

\bibitem{KTW} T.R.~Kirkpatrick and D.~Thirumalai,
  Phys.~Rev.~Lett. {\bf 58}, 2091 (1987); Phys. Rev. B {\bf 36}, 5388
  (1987); T.R. Kirkpatrick, D. Thirumalai and P.G. Wolynes, Phys.
  Rev. A {\bf 40}, 1045 (1989).

\bibitem{MCT} W. Gotze, {\it Complex Dynamics of Glass-Forming
  Liquids} (Oxford University Press, Oxford, 2009).

\bibitem{BBRFOT} G. Biroli and J.-P. Bouchaud, {\it The Random
  First-Order Transition Theory of Glasses: a critical assessment},
  arXiv:0912.2542 (2009).

\bibitem{FPchi4} S. Franz and G. Parisi, J. Phys. Cond. Mat. {\bf
  12}, 6335 (2000). C. Donati, S. Franz, G. Parisi and S.C. Glotzer,
  J. of Non-Cryst. Solids {\bf 307-310}, 215 (2002).

\bibitem{BB} J. Bouchaud and G. Biroli, Europhys. Lett. {\bf 67}, 21
  (2004).

\bibitem{FP} S. Franz, G. Parisi, J. Phys. I (France) {\bf 5}, 1401
  (1995); Phys. Rev.  Lett. {\bf 79} 2486 (1997); Physica A {\bf 261},
  317 (1998).

\bibitem{Monasson} R. Monasson, Phys. Rev. Lett. {\bf 75}, 2847
  (1995).

\bibitem{MP} M. Mezard and G. Parisi, {\it Replicas and Glasses},
  arXiv:0910.2838 (2009).

\bibitem{SZ} G. Szamel, Europhys. Lett. {\bf 91}, 56004 (2010).

\bibitem{TDP} T. Temesvari, C. De Dominicis and I. R. Pimentel,
  Eur. Phys. J. B {\bf 25}, 361 (2002).

\bibitem{DDG} C. De Dominicis and I. Giardina, {\it Random Fields and
  Spin Glasses} (Cambridge University Press, Cambridge, 2006).

\bibitem{PS1} G. Parisi and N. Sourlas, Phys. Rev. Lett. {\bf 43}, 744
  (1979).

\bibitem{Cardy} J.L. Cardy, Phys. Lett. B {\bf 125}, 470 (1983);
  Physica D {\bf 15}, 123 (1985).

\bibitem{PS2} G. Parisi and N. Sourlas, Phys. Rev. Lett. {\bf 46}, 871
  (1981).

\bibitem{SBBB} T. Sarlat, A. Billoire, G. Biroli and J.-P. Bouchaud,
  J. Stat. Mech. P08014 (2009).

\bibitem{NishiInit} Y. Ozeki, J. Phys. A: Math. Gen. {\bf 28}, 3645
  (1995). F. Krzakala and L. Zdeborova, J. Chem. Phys. {\bf 134},
  034513 (2011).

\bibitem{MRT} A. Montanari and F. Ricci-Tersenghi, Eur. Phys. J. B
  {\bf 33}, 339 (2003); Phys. Rev. B {\bf 70}, 134406 (2004).

\bibitem{HKproc} S. Franz, G. Parisi, F. Ricci-Tersenghi and T. Rizzo,
  preprint arxiv:1008:0996 (2010).

\end{thebibliography}
\end{document}